\documentclass[aps,prc,twocolumn,nofootinbib,preprintnumbers,amsmath,amssymb]{revtex4}

\usepackage{graphicx}
\usepackage{dcolumn}
\usepackage{bm}
\usepackage{epsfig}
\usepackage{longtable}

\def\fun#1#2{\lower3.6pt\vbox{\baselineskip0pt\lineskip.9pt
\ialign{$\mathsurround=0pt#1\hfil##\hfil$\crcr#2\crcr\sim\crcr}}}

\begin{document}

\preprint{}

\title{
Precise measurement of $\alpha_K$ for the 88.2-keV $M$4 transition in $^{127}$Te: Test of internal-conversion theory
}

\author{N. Nica}
\email{nica@comp.tamu.edu}

\author{J.C. Hardy}
\email{hardy@comp.tamu.edu}

\author{V.E. Iacob}

\author{H.I. Park}

\author{K. Brandenburg}
\altaffiliation {REU summer student from Clemson University, Clemson, SC}

\affiliation{ Cyclotron Institute, Texas A\&M University, College Station, Texas 77843, USA}
\homepage{http://cyclotron.tamu.edu/}

\author{M.B. Trzhaskovskaya}
\affiliation{Petersburg Nuclear Physics Institute, Gatchina 188300, Russia}

\date{\today}

\begin{abstract}
We have measured the $K$-shell internal conversion coefficient, $\alpha_K$, for the 88.2-keV
$M$4 transition from the 106-day isomer to the ground state in $^{127}$Te to be 484(6).  When compared with Dirac-Fock calculations of 
$\alpha_K$, this result agrees well with the version of the theory that incorporates the effect of the
$K$-shell atomic vacancy and disagrees with the one that does not.  As a byproduct of this measurement,
we have determined the beta branching from the isomer to be 2.14(3)\%.
 
\end{abstract}

\maketitle

\section{\label{sec:introd} INTRODUCTION}

Starting in 2004, we have been publishing a sequence of papers \cite{Ni04,Ni05,Ni07,Ni08,Ni09,Ni14,Ha14,Ni16}, in which we report 
measurements of $K$-shell Internal Conversion Coefficients (ICCs) for $E$3 and $M$4 transitions with a precision
of $\pm$2\% or better. The motivation has been to test ICC theory, in particular its treatment of the $K$-shell vacancy left
behind by the emitted electron.  When we began, the then-current survey of world data on ICCs \cite{Ra02} included only
six $\alpha_K$ values for high-multipolarity transitions ($E$3, $M$3 and above) measured to two-percent precision or better. At
the time, the surveyors concluded that the data favored the ICC calculation that ignored the atomic vacancy.  Combining our
work since then with the result reported here, we have now nearly doubled the number of precisely measured $\alpha_K$ values and
have changed that conclusion.

What makes such precise measurements possible for us is our having an HPGe detector whose relative efficiency is known to
$\pm$0.15\% ($\pm$0.20\% absolute) over a wide range of energies: See, for example, Ref.\,\cite{He03}.  By detecting both the
$K$ x rays and the $\gamma$ rays from a transition of interest in the same well-calibrated detector at the same time, we can
avoid many sources of error.

By 2008, our early results from this program influenced a reevaluation of ICCs by Kib\'{e}di $et~al.$ \cite{Ki08}, who also
developed BrIcc, a new data-base obtained from the basic code by Band $et~al.$ \cite{Ba02}.  In conformity with our
conclusions, BrIcc employed a version of the code that incorporates the vacancy in the ``frozen orbital" approximation.  The
BrIcc data-base has been adopted by the National Nuclear Data Center (NNDC) and is available on-line for the determination
of ICCs.  Our experimental results obtained since 2008 continue to support that decision and now include transitions in
seven nuclei that cover the range $48 \leq Z \leq 78$.

The measurement we report here is of the 88.23-keV, $M$4 transition from the 106-day isomeric state in $^{127}$Te to its
ground state.  The calculation of $\alpha_K$ for this transition depends appreciably on whether the atomic vacancy is
accounted for, there being a 3.7\% difference between the values obtained with and without inclusion of the vacancy. The
only previous reported measurement of the $\alpha_K$ value \cite{So77} had an uncertainty of $\pm$5\%, which overlaps
both calculated values. Our new result has $\pm$1.2\% precision and clearly distinguishes between the two options.

\section {\label{overview} Measurement Overview}

We have described our measurement techniques in detail in previous publications \cite{Ni04,Ni07}
so only a summary will be given here.  If a decay scheme is dominated by a single transition
that can convert in the atomic $K$ shell, and a spectrum of $K$ x rays and $\gamma$ rays is recorded
for its decay, then the $K$-shell internal conversion coefficient for that transition is given by
\begin{equation}
\alpha_K \omega_K = \frac{N_K}{N_\gamma} \cdot \frac{\epsilon_\gamma}{\epsilon_K},
\label{alpha}
\end{equation}
where $\omega_K$ is the fluorescence yield; $N_K$ and $N_{\gamma}$ are the total numbers of observed
$K$ x rays and $\gamma$ rays, respectively; and $\epsilon_K$ and $\epsilon_\gamma$ are the
corresponding photopeak detection efficiencies.

The fluorescence yield for tellurium has been measured several times, with a weighted average quoted to $\pm$3.1\% \cite{Ka12}.
Furthermore, world data for fluorescence yields have also been evaluated systematically as a function of $Z$ \cite{Sc96}
for all elements with $10 \leq Z \leq 100$, and $\omega_K$ values have been recommended for each element in this range.  The
recommended value for tellurium, $Z$ = 52, is 0.875(4), which is consistent with the average measured value but has a smaller
relative uncertainty, $\pm$0.5\%.  We use this value.

The decay scheme of the 106-day isomer in $^{127}$Te is shown in Fig.\,\ref{fig1}. With a single electromagnetic decay path,
directly feeding the ground state, it clearly satisfies the condition required for the validity of Eq.\,(\ref{alpha}).  The
only complication is that the isomer also has a 2.1\% $\beta$-branch to an excited state in $^{127}$I, which decays by a 57.6-keV
transition with an $\alpha_K$ value of 3.16 \cite{Ha11}.  This leads unavoidably to the presence of iodine $K$ x rays.  Though their
intensity is only a few percent that of the predominant tellurium x rays, the two groups are unresolved from one another in our detector,
so the iodine component must be carefully accounted for.  The $^{127}$Te ground state $\beta$ decays as well, but only
weakly populates an excited state, the decay of which produces considerably less conversion. 

In our experiment, the HPGe detector we used to observe both $\gamma$ rays and $K$ x rays has been meticulously calibrated
\cite{Ha02,He03,He04} for efficiency to sub-percent precision, originally over an energy range from 50 to 3500 keV but
more recently extended \cite{Ni14} with $\pm$1\% precision down to 22.6 keV, the average energy of silver $K$ x rays.  Over this whole
energy region, precise measured data were combined with Monte Carlo calculations from the CYLTRAN code \cite{Ha92} to yield a very precise
and accurate detector efficiency curve.  In our present study, the $\gamma$ ray of interest at 88.2 keV is well within the energy region for
which our efficiencies are known to a relative precision of $\pm$0.15\%.  The tellurium $K$ x rays lie between 27 and 32 keV, comfortably
within our extended region of calibration, so the detector efficiency for them can be quoted to a relative precision of $\pm$1\%.

\section{\label{exp} Experiment}

\subsection{\label{sprep} Source Preparation}

We obtained tellurium powder enriched to $\geq$98\% in $^{126}$Te from Trace Sciences International Corp.  As delivered, the grain size was  25$\pm$5\,$\mu$m,
but by grinding it between glass slides we reduced it to a relatively uniform grain size of about 1\,$\mu$m as determined with a microscope.  A
small quantity of this powder was sprinkled over a 1-cm diameter disk of adhesive Mylar, 25-$\mu$m thick, and another layer of the same adhesive
Mylar was placed on top, sealing the powder between.  A second ``dummy'' sample, identical except for the absence of any tellurium, was prepared too.

The tellurium and dummy samples were activated together for 24\,h in a neutron flux of $\sim7.5\times10^{12}\,n$/(cm$^2$\,s) at the TRIGA reactor
in the Texas A\&M Nuclear Science Center.  After removal from the reactor, both samples were found to have curled up.  As we carefully unrolled it,
the tellurium sample cracked, requiring us to secure it between two fresh Mylar sheets, 12.5-$\mu$m thick. The final source thus comprised
a 4$\pm$2\,-$\mu$m-thick layer of tellurium enclosed between layers of Mylar 37.5-$\mu$m thick.  We then stored the samples for one month to
allow short-lived activities to die out. 

\begin{figure}[t]
\epsfig{file=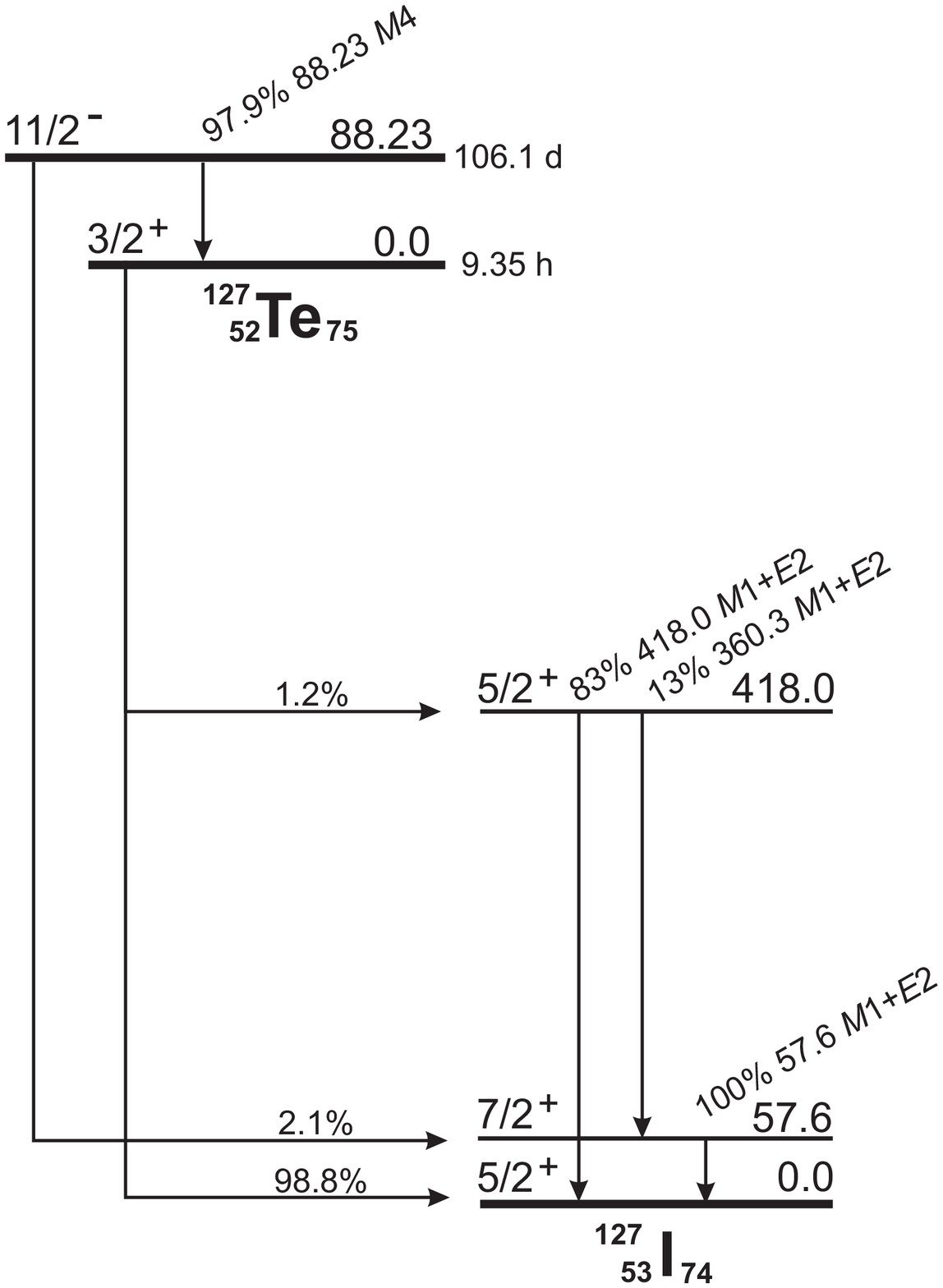,width=5.2cm}
\caption{Decay scheme for the 106-day isomer in $^{127}$Te, illustrating the channels important to this measurement.  The data are taken from Ref.\,\cite{Ha11} except
for the $\beta$ branching ratio from the isomeric state and the $\gamma$ branching of the 418.0 keV level, which reflect new results from this work. For clarity, a
weak (4\%) electromagnetic branch from the 418.0-keV state has been omitted.}
\label{fig1}
\end{figure}

\subsection{\label{decaymeas} Radioactive decay measurements}

We acquired spectra with our precisely calibrated HPGe detector and with the same electronics used in its calibration \cite{He03}.  Our
analog-to-digital converter was an Ortec TRUMP$^{TM}$-8k/2k card controlled by MAESTRO$^{TM}$ software.  We acquired 8k-channel spectra at a
source-to-detector distance of 151~mm, the distance at which our calibration is well established.  Each spectrum covered the energy interval
10-2000 keV with a dispersion of about 0.25 keV/channel.

After energy-calibrating our system with a $^{152}$Eu source, we recorded sequential decay spectra from the tellurium sample in two one-week
periods.  Before after and between these two periods, we interspersed measurements of the dummy sample and of room background.  These were
used to aid us in identifying the origin of all peaks in the spectrum.

\section{\label{sec:analysis} Analysis}
\subsection{\label{subsec:peakfit} Peak fitting}

\begin{figure*}[t]
\epsfig{file=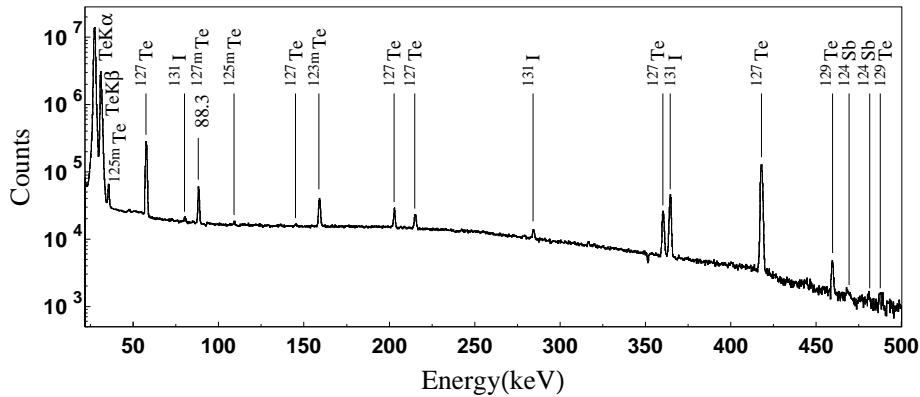,width=13cm}
\caption{Portion of the background-subtracted x- and $\gamma$-ray energy spectrum recorded over a period of two weeks, more than a month after
activation of enriched $^{126}$Te.  Peaks are labeled by their $\beta$-decay parent.  The inverted ``peak'' at 352 keV results from a subtracted
background $\gamma$ ray present in the decay of $^{214}$Pb, which is a granddaughter of environmental $^{222}$Rn and consequently is not constant in time. }
\label{fig2}
\end{figure*}

A portion of the background-subtracted spectrum recorded from the tellurium source is presented in Fig.\,\ref{fig2}: It includes the x- and
$\gamma$-ray peaks of interest from the decay of $^{127m}$Te, as well as a number of peaks from contaminant activities.  In our analysis
of the data, we followed the same methodology as we did with previous source measurements \cite{Ni04,Ni05,Ni07,Ni08,Ni09,Ni14,Ha14, Ni16}.  
We first extracted areas, not only for the $^{127m}$Te peaks, but also for essentially all the other x- and $\gamma$-ray peaks in the spectrum.  Our
procedure was to determine the areas with GF3, the least-squares peak-fitting program in the RADWARE series \cite{Rapc}.  In doing so, we used
the same fitting procedures as were used in the original detector-efficiency calibration \cite{Ha02,He03,He04}.

\begin{figure}[b]
\epsfig{file=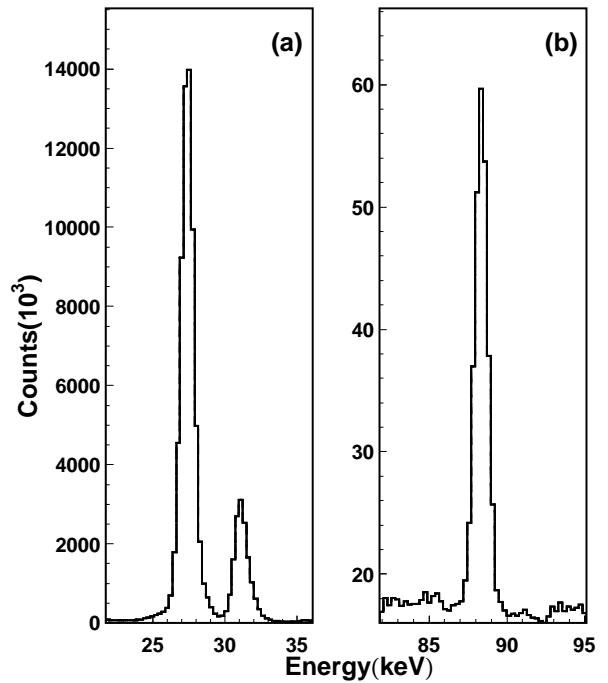,width=8.5cm}
\caption{Spectra for the two energy regions of interest in this measurement, the one on the left including the tellurium $K$ x rays and the one
on the right, the $\gamma$-ray peak at 88.2 keV. These correspond to the full spectrum presented in Fig.\,\ref{fig2}}
\label{fig3}
\end{figure}

Figure \ref{fig3} shows expanded versions of the two energy regions of interest for this measurement: one encompassing the tellurium $K$ x
rays and the other, the $\gamma$ ray at 88.2 keV.  The tellurium x-ray peaks, which of course include some unresolved impurity x rays, lie cleanly
on a low flat background; the $\gamma$ ray is much weaker relative to the local background but is nevertheless easily analyzed.

In what follows we do not distinguish between $K_{\alpha}$ and $K_{\beta}$ x rays.  Scattering effects are quite pronounced at these energies and they
are difficult to account for with an HPGe detector when peaks are close together, so we have chosen as before to use only the sum of the $K_{\alpha}$
and $K_{\beta}$ x-ray peaks.  For calibration purposes, we consider each sum to be located at the intensity-weighted average energy of the component
peaks\footnote[1]{To establish the weighting, we used the intensities of the individual x-ray components from Table 7a in Ref. \cite{Fi96}.}---28.03
keV for tellurium and 29.20 keV for iodine.

The total number of counts recorded in the combined x-ray peaks and in the 88.2-keV $\gamma$-ray peak must next be corrected for impurities and
for other effects.  These corrections are described in the following sections.

\subsection{\label{subsec:imp} Impurities}

Once the areas (and energies) of all $\gamma$-ray peaks had been established, we could identify every impurity in the $^{127m}$Te spectrum and carefully check
to see if any were known to produce x or $\gamma$ rays that might interfere with the tellurium $K$ x rays or the $\gamma$-ray peak of interest at
88.2 keV.  As is evident from Fig.\,\ref{fig2}, even the weakest peaks were identified.  In all, we found 10 contaminant activities that
contribute in some way to the tellurium x-ray region; these are listed in Table\,\ref{table1}, where the contributions are given as percentages
of the total x rays recorded.  Only $^{125m}$Te contributes more than a percent; most of the rest are extremely weak. As to the $\gamma$-ray
peak, only one impurity, $^{123m}$Te, interferes with it in any way and the impact is very slight. 

The count totals recorded for the combined $K$ x-ray peaks and for the 88.2-keV peak both appear in Table\,\ref{table2}, with their respective impurity
components listed immediately below them.  The impurity totals correspond to the percentage breakdowns given in Table\,\ref{table1}.

\begin{table}[t]
\caption{\label{table1} The contributions of identified impurities to the energy regions of the tellurium $K$ x-ray peaks and the 88.2-keV $\gamma$-ray peak.}
\vspace{2mm}
\begin{ruledtabular}
\begin{tabular}{lll}
 & & Contaminant \\
Source & Contaminant &  contribution (\%) \\[1mm]
\hline \\[-2mm]
& &   \\[-4mm]
To $K$ x-ray peaks & & \\[1mm]
~~~$^{121}$Te & Sb $K$ x rays & 0.0162(12) \\
~~~$^{110m}$Ag & Cd+Ag $K$ x rays & 0.000458(10) \\
~~~$^{121m}$Te & Te+Sb $K$ x rays & 0.0022(3)   \\
~~~$^{123m}$Te & Te $K$ x rays & 0.095(3)  \\
~~~$^{124}$Sb & Te $K$ x rays & 0.0135(3)   \\
~~~$^{125m}$Te & Te $K$ x rays & 2.77(8)  \\
~~~$^{125m}$Te & 35.5-keV $\gamma$ ray & 0.176(2)  \\
~~~$^{129}$Te & 27.8-keV $\gamma$ ray & 0.11(4)  \\
~~~$^{129m}$Te & Te $K$ x rays &   \\
 & +27.8-keV $\gamma$ ray & 0.289(11)  \\
~~~$^{131}$I & Xe $K$ x rays & 0.0346(8)  \\
~~~$^{131m}$Xe & Xe $K$ x rays & 0.005(5)  \\[1mm]
To 88.2-keV peak & & \\[1mm]
~~~$^{123m}$Te & 88.5-keV $\gamma$ ray & 0.078(2)  \\
\end{tabular}
\end{ruledtabular}
\end{table}
\subsection{\label{subsec:eff} Efficiency ratios}

In order to determine $\alpha_K$ for the 88.2-keV $M$4 transition in $^{127}$Te, we require the efficiency ratio, $\epsilon_{\gamma 88.2}/\epsilon_{K28.0}$
as can be seen in Eq.\,(\ref{alpha}).  Following the same procedure as the one we used in analyzing the decay of $^{119m}$Sn \cite{Ni14}, we
employ as low-energy calibration the well-known decay of $^{109}$Cd, which emits 88.0-keV $\gamma$ rays and silver $K$ x rays at a weighted average
energy of 22.57 keV.  Both are close in energy to the respective $\gamma$ and x rays observed in the current measurement.  

We obtain the required ratio, $\epsilon_{\gamma 88.2}/\epsilon_{K28.0}$ from the following relation:
\begin{equation}
\frac{\epsilon_{\gamma 88.2}}{\epsilon_{K28.0}} = \frac{\epsilon_{\gamma88.0}}{\epsilon_{K22.6}} \cdot
 \frac{\epsilon_{\gamma 88.2}}{\epsilon_{\gamma 88.0}} \cdot \frac{\epsilon_{K22.6}}{\epsilon_{K28.0}}.
\label{effratio}
\end{equation} 
We take the $^{109}$Cd ratio $\epsilon_{\gamma 88.0}/\epsilon_{K22.6}$ from our previously reported measurement \cite{Ni14}.  The ratio
$\epsilon_{\gamma 88.2}/\epsilon_{\gamma 88.0}$ is very nearly unity and determined with high precision from our known detector efficiency curve
calculated with the CYLTRAN code \cite{He03}, while $\epsilon_{K22.6}/\epsilon_{K28.0}$ comes from a CYLTRAN calculation as well but in an energy region
with higher relative uncertainty.  Nevertheless, the energy span is not large so the uncertainty is only $\pm$0.5\%.  The values of all four efficiency
ratios from Eq.\,(\ref{effratio}) appear in the third block of Table \ref{table2}.

\subsection{\label{subsec:beta} Beta decay contribution}

Although the decay of $^{127m}$Te occurs predominantly through the 88.2-keV $M$4 electromagnetic transition, it also has a weak $\beta$-decay branch that
populates the 57.6-keV level in $^{127}$I.  That level is additionally populated by another weak $\beta$-decay branch from the ground state of $^{127}$Te,
the decay of which is in secular equilibrium with the isomeric-state decay.  This is illustrated in Fig.\,\ref{fig1}.  The transition from the 57.6-keV state
to the ground state of $^{127}$I has $M$1+$E$2 character with a measured mixing ratio and an $\alpha_{K57.6}$ value of 3.16(5) \cite{Ha11}. This
gives rise to iodine $K$ x rays, which are unresolved from the tellurium x rays and thus must be corrected for. 

If $N_{\gamma 57.6}$ is the total number of counts in the 57.6-keV peak, then the corresponding number of iodine $K$ x rays that appear in the spectrum,
$N_{K29.2}$, is given by
\begin{equation}
N_{K29.2} = N_{\gamma 57.6} \cdot \frac{\epsilon_{K29.2}}{\epsilon_{\gamma 57.6}} \cdot \alpha_{K57.6} \cdot \omega_{Ki} ,
\label{bdecaycomp}
\end{equation}
where $\omega_{Ki}$ = 0.882(4) is the fluorescence yield for iodine \cite{Sc96}.  

Using the same approach we took for Eq.\,(\ref{effratio}), we obtain the required efficiency ratio, $\epsilon_{K29.2}/\epsilon_{\gamma 57.6}$, from the following relation:
\begin{equation}
\frac{\epsilon_{K29.2}}{\epsilon_{\gamma 57.6}} = \frac{\epsilon_{K22.6}}{\epsilon_{\gamma88.0}} \cdot
 \frac{\epsilon_{\gamma 88.0}}{\epsilon_{\gamma 57.6}} \cdot \frac{\epsilon_{K29.2}}{\epsilon_{K22.6}},
\label{effratioI}
\end{equation} 
in which our previously measured $^{109}$Cd ratio \cite{Ni14} again plays a key role. Here it is inverted compared to Eq.\,(\ref{effratio}) and has the value
$\epsilon_{K22.6}/\epsilon_{\gamma 88.0}$ = 0.935(7).   Using our known detector efficiency curve calculated with the CYLTRAN code, we determine
the other two ratios to be $\epsilon_{\gamma 88.0}/\epsilon_{\gamma 57.6}$ = 0.9796(11) and $\epsilon_{K29.2}/\epsilon_{K22.6}$ = 1.073(5).  The
final result for $\epsilon_{K29.2}/\epsilon_{\gamma 57.6}$ thus becomes 0.983(9).

Substituting this result into Eq.\,(\ref{bdecaycomp}), together with the measured number of counts in the 57.6-keV $\gamma$-ray peak, $N_{\gamma 57.6} =
1.0146(15) \times 10^6$, we obtain the contribution of the $\beta$-decay channels to the x-ray spectrum to be $N_{K 29.2} = 2.78(5)\times10^6$.  This result
also appears in the first block of Table\,\ref{table2}.

\begin{table}[t]
\caption{\label{table2}Corrections to the $^{127}$Te $K$ x rays and the 88.2-keV $\gamma$ ray as well as the additional
information required to extract a value for $\alpha_K$. }
\vspace{2mm}
\begin{ruledtabular}
\begin{tabular}{lll}
Quantity   &  Value  & Source  \\
\hline \\[-2mm]
\multicolumn{3}{l}{Te ($K_{\alpha} + K_{\beta}$) x rays}  \\
~~ Total counts & 7.8923(12) $\times 10^7$ & Sec.~\ref{subsec:peakfit}  \\
~~ Impurities  & -2.77(7)$\times 10^6$  & Sec.~\ref{subsec:imp}  \\
~~ $\beta$-decay contribution  & -2.78(5)$\times 10^6$   &  Sec.~\ref{subsec:beta} \\
~~ Lorentzian correction  &  +0.12(2)\%  &  Sec.~\ref{subsec:Lor} \\
~~ Net corrected counts, $N_{K28.0}$  & 7.346(9)$\times 10^7$  &  \\
\hline \\[-2mm]
\multicolumn{3}{l}{$^{127}$Te 88.2-keV $\gamma$ ray}  \\
~~ Total counts & 1.760(13)$\times 10^5$ & Sec.~\ref{subsec:peakfit}  \\
~~ Impurities  &  -1.38(4)$\times 10^2$  &  Sec.~\ref{subsec:imp}  \\
~~ Net corrected counts, $N_{\gamma\,88.2}$ & 1.759(13)$\times 10^5$ &  \\
\hline \\[-2mm]
\multicolumn{3}{l}{Efficiency calculation}  \\
~~ $\epsilon_{\gamma\,88.0}$/$\epsilon_{K22.6}$ & 1.069(8) & \cite{Ni14} \\
~~ $\epsilon_{\gamma\,88.2}$/$\epsilon_{\gamma\,88.0}$ & 1.0011(1) & \cite{He03} \\
~~ $\epsilon_{K22.6}$/$\epsilon_{K28.0}$ & 0.940(5) & \cite{He03} \\
~~ $\epsilon_{\gamma\,88.2}$/$\epsilon_{K28.0}$ & 1.006(9) & \\
\hline \\[-2mm]
\multicolumn{3}{l}{Evaluation of $\alpha_K$}  \\
~~ $N_{K28.0}/N_{\gamma\,88.2}$  &  418(3)  & This table  \\
~~ Relative attenuation  &  +0.9(4)\%  &  Sec.\,\ref{subsec:att} \\
~~ $\omega_K$  &  0.875(4)  &  \cite{Sc96}  \\
~~ $\alpha_K$ for 88.2-keV transition  & 484(6)   &  Eq.\,\ref{alpha} \\
\vspace{-10.pt}
\end{tabular}
\end{ruledtabular}
\end{table}

\subsection{\label{subsec:Lor} Lorentzian correction}

As explained in our previous papers (see, for example, Ref.\,\cite{Ni04}) we use a special modification of the GF3 program
that allows us to sum the total counts above background within selected energy limits.  To account for possible missed counts outside those limits, the
program adds an extrapolated Gaussian tail.  This extrapolated tail does not do full justice to x-ray peaks, whose Lorentzian shapes reflect the finite
widths of the atomic levels responsible for them.  To correct for this effect we compute simulated spectra using realistic Voigt functions to generate
the x-ray peaks, and we then analyze them with GF3, following exactly the same fitting procedure as is used for the real data, to ascertain how
much was missed by this approach.  The resultant correction factor appears as a percent in the first block of Table\,\ref{table2}.

\subsection{\label{subsec:att} Attenuation in the sample}

As described in Sec.~\ref{sprep}, our source was a thin layer of tellurium enclosed between Mylar sheets.  We obtained the attenuation both of the tellurium
x rays and of the 88.2-keV $\gamma$ ray using standard tables of attenuation coefficients \cite{Ch05}.  Since we are striving to evaluate $\alpha_K$ from
Eq.\,(\ref{alpha}), what matters in that context is the attenuation for the x rays relative to that for the 88.2-keV $\gamma$ ray.  We determine that the x
rays suffered 0.9(4)\% more attenuation and it is this number that appears in the bottom block of Table\,\ref{table2}.

\section{\label{sec:results} Results and Discussion}

\subsection{\label{subsect:ICC} $\alpha_K$ for the 88.2-keV transition}

\begin{table}[b]
\caption{\label{table3}Comparison of the measured $\alpha_K$ values for the 88.23(7)-keV $M$4 transition from $^{127m}$Tm with calculated values based on
three different theoretical models for dealing with the $K$-shell vacancy.  Shown also are the percentage deviations, $\Delta$, from the experimental value
calculated as (experiment-theory)/theory.  For a description of the various models used to determine the conversion coefficients, see text and Ref.\,\cite{Ni04}.}
\vspace{2mm}
\begin{ruledtabular}
\begin{tabular}{lll}
\multicolumn{1}{l}{Model}  & \multicolumn{1}{c}{~~$\alpha_K$} & \multicolumn{1}{c}{~~$\Delta$(\%)}  \\
\hline \\[-3mm]
Experiment & 484(6)  &  \\
Theory: & & \\
~~~No vacancy  & 468.6(17) & +3.3(13)  \\
~~~Vacancy, frozen orbitals  & 486.4(17) & $-0.5(13)$  \\
~~~Vacancy, SCF of ion  & 483.1(17) & $+0.2(13)$  \\
\vspace{-10.pt}
\end{tabular}
\end{ruledtabular}
\end{table}

The third and fourth blocks of Table\,\ref{table2} contain all the data required to determine the value of $\alpha_K$ for the 88.2-keV transition from
Eq.\,(\ref{alpha}).  The result, $\alpha_K$ = 484(6), agrees very closely with the only previous measurement  \cite{So77} of this ICC, 484(23), but has replaced
a value having $\pm$5\% precision with one having $\pm$1.2\%. 

Our measured $\alpha_K$ value is compared with three different theoretical calculations in Table\,\ref{table3}. All three calculations were made within
the Dirac-Fock framework, but one ignores the presence of the $K$-shell vacancy while the other two include it using different approximations: the frozen
orbital approximation, in which it is assumed that the atomic orbitals have no time to rearrange after the electron's removal; and the SCF approximation, in
which the final-state continuum wave function is calculated in the self-consistent field (SCF) of the ion, assuming full relaxation of the ion orbitals.  To
obtain these results we used the value 88.23(7) keV \cite{Ha11} for the $^{127m}$Te transition energy.  The experimental uncertainty in this number is reflected
in the uncertainties quoted on the theoretical values of $\alpha_K$ in the table.

The percentage deviations given in Table\,\ref{table3} indicate excellent agreement between our measured result and the two calculations that include some
provision for the atomic vacancy.  Our measurement disagrees by 2.5 standard deviations with the calculation that ignores the vacancy.  This outcome is consistent
with our previous six precise $\alpha_K$ measurements on $E$3 and $M$4 transitions in $^{111}$Cd \cite{Ni16}, $^{119}$Sn \cite{Ni14,Ha14}, $^{134}$Cs
\cite{Ni07,Ni08}, $^{137}$Ba \cite{Ni07,Ni08}, $^{193}$Ir \cite{Ni04,Ni05} and $^{197}$Pt \cite{Ni09}, all of which agreed well with calculations that included
the vacancy, and disagreed -- some by many standard deviations -- with the no-vacancy calculations.   

\subsection{\label{subsect:betabr} $\beta$-decay branching ratio from $^{127m}$Te}

As remarked in Sec.\,\ref{subsec:beta} and illustrated in Fig.\,\ref{fig1}, the 57.6-keV level in $^{127}$I is populated by two $\beta$-decay branches, one from
the isomeric state and the other from the ground state of $^{127}$Te.  Since the ground state has a half-life of 9.35 hours and our spectrum was acquired more than a
month after activation, the two decays were certainly in secular equilibrium for our measurement.  Under these conditions, we see from the decay
scheme that the 57.6-keV state is fed $\sim$14 times more strongly from the isomer than it is from the ground state.  As a consequence, it is possible to
use the ratio of intensities of the $\gamma$-ray peaks at 57.6 and 88.2 keV to extract a rather precise value for the strength of the stronger $\beta$ branch even
though the strength of the weaker branch remains relatively imprecise.

If we represent the $\beta$ branching ratio from the isomer by $R$, then we can write:
\begin{equation}
\frac{R}{1-R} = \frac{N_{\gamma 57.6}}{N_{\gamma 88.2}}\cdot\frac{\epsilon_{\gamma88.2}}{\epsilon_{\gamma57.6}}\cdot\frac{(1+\alpha_{T57.6})}{(1+\alpha_{T88.2})} - 0.00155(21),
\label{R}
\end{equation} 
where the $\alpha_T$ values are the total ICCs for the indicated transitions, $\alpha_{T57.6}$ = 3.72(3) and $\alpha_{T88.2}$ = 1138(5), for which we have used the
Dirac-Fock ``frozen orbital'' calculated values.  The numerical term, 0.00155(21) is the product of the $\beta$ branching ratio for the transition from the ground
state of $^{127}$Te to the 418.0-keV state in $^{127}$I, and the branching ratio for the subsequent electromagnetic decay of the 418.0-keV state to populate the 57.6-keV
state.  We take the former, 0.0119(16), from Ref\,\cite{Ha11}, while the latter we obtain from our own $\gamma$-ray spectrum.

\begin{table}[t]
\caption{\label{table4}Relative intensities of $\gamma$ rays de-exciting the 418.0-keV level in $^{127}$I.}
\vspace{2mm}
\begin{ruledtabular}
\begin{tabular}{llll}
\multicolumn{1}{c}{Energy}  & \multicolumn{3}{c}{Relative $\gamma$-ray intensity}  \\
\cline{2-4}  \\[-3mm]
\multicolumn{1}{l}{\,(keV)}  & \multicolumn{1}{l}{Ref.\,\cite{Au65}} & \multicolumn{1}{l}{Ref.\,\cite{Ap70}} & \multicolumn{1}{l}{This work} \\
\hline \\[-3mm]
~418.0 & 100 & 100 & 100  \\
~360.3 & 14.8(1) & 13.6(4) & $  15.2(1) $  \\
~215.2 & 3.9(2) & 3.91(17) & $  4.44(7)$  \\
\vspace{-10.pt}
\end{tabular}
\end{ruledtabular}
\end{table}

Table\,\ref{table4} lists the relative intensities we measure for the three $\gamma$ rays that de-excite the 418-keV state in $^{127}$I. The two previous
measurements, with which our results are compared, were published a half century ago, with now-untraceable efficiency calibrations and likely rather
optimistic uncertainties.  We consider our results to be more reliable and use them exclusively.  After adjusting for internal conversion, we determine the
electromagnetic branching ratio for the 360.3-keV transition, which feeds the 57.6-keV state, to be 0.128(1).  In addition, the 215.2-keV transition has a small
probability for populating the 57.6-keV state \cite{Ha11}, so the total feeding of the 57.6-keV state from the decay of the 418.0-keV state becomes 0.130(1).
We use this result in determining the numerical term in Eq.\,(\ref{R}).

Solving Eq.\,(\ref{R}) for $R$, we determine the $\beta$ branching of the isomeric state, $^{127m}$Te, to be 2.14(3)\%.  This compares favorably with, but is 7 times
more precise than, the only previous measurement of this quantity, 2.4(2)\%, which was published in 1970 \cite{Ap70}.

\vspace{8mm}

\section{Conclusions}
Our measurement of the $K$-shell internal conversion coefficient for the 88.2-keV $M$4 transition from $^{127m}$Te has yielded a value, $\alpha_K$ = 484(6), which
agrees with a version of the Dirac-Fock theory that includes the atomic vacancy.  It disagrees (by $\sim$2.5$\sigma$) with theory if the vacancy is ignored.  
We have now made seven precise $\alpha_K$ measurements for $E$3 and $M$4 transitions in nuclei with a wide range of $Z$ values.  Their corresponding
conversion-electron energies also ranged widely, from $\sim$4 keV in $^{193}$Ir to $\sim$630 keV in $^{137}$Ba.  These measurements together present a consistent
pattern that supports the Dirac-Fock theory for calculating internal conversion coefficients provided that it takes account of the atomic vacancy.

As a byproduct of the current measurement, we have also improved considerably the $\beta$-branching ratio for the 88.2-keV isomeric state in $^{127}$Te.

\begin{acknowledgments}

We thank the Texas A\&M Nuclear Science Center staff for their help with the neutron
activations.  This material is based upon work supported by the U.S. Department of Energy, Office of Science,
Office of Nuclear Physics, under Award Number DE-FG03-93ER40773, and by the Robert A. Welch Foundation under
Grant No.\,A-1397.

\end{acknowledgments}


\begin{thebibliography}{}

\bibitem{Ni04} 
N. Nica, J. C. Hardy, V. E. Iacob, S. Raman, C. W. Nestor Jr., and M. B. Trzhaskovskaya, 
\prc{\bf 70}, 054305 (2004).

\bibitem{Ni05}
N. Nica, J. C. Hardy, V. E. Iacob, J. R. Montague, and M. B. Trzhaskovskaya, 
\prc{\bf 71}, 054320 (2005).

\bibitem{Ni07}
N. Nica, J. C. Hardy, V. E. Iacob, W. E. Rockwell, and M. B. Trzhaskovskaya, 
\prc{\bf 75}, 024308 (2007).

\bibitem{Ni08}
N. Nica, J. C. Hardy, V. E. Iacob, C. Balonek, and M. B. Trzhaskovskaya, 
\prc{\bf 77}, 034306 (2008).

\bibitem{Ni09}
N. Nica, J. C. Hardy, V. E. Iacob, J. Goodwin, C. Balonek, M. Hernberg, J. Nolan and M. B. Trzhaskovskaya, 
\prc{\bf 80}, 064314 (2009).

\bibitem{Ni14}
N. Nica, J. C. Hardy, V. E. Iacob, M. Bencomo, V. Horvat, H.I. Park, M. Maguire, S. Miller and M. B. Trzhaskovskaya, 
\prc{\bf 89}, 014303 (2014).

\bibitem{Ha14}
J. C. Hardy, N. Nica, V. E. Iacob, S. Miller, M. Maguire and M. B. Trzhaskovskaya
Appl. Rad and Isot. {\bf 87}, 87 (2014).

\bibitem{Ni16}
N. Nica, J. C. Hardy, V. E. Iacob, T.A. Werke, C.M. Folden III, L. Pineda and M. B. Trzhaskovskaya, 
\prc{\bf 93}, 034305 (2016).

\bibitem{Ra02} 
S. Raman, C. W. Nestor, Jr., A. Ichihara, and M. B. Trzhaskovskaya, 
\prc{\bf 66}, 044312 (2002); see also the electronic addendum to this paper, the
location of which is given in the paper's reference 32.

\bibitem{He03}
R. G. Helmer, J. C. Hardy, V. E. Iacob, M. Sanchez-Vega, R. G. Neilson, and J. Nelson,
Nucl. Instrum. Methods Phys. Res. A {\bf 511}, 360 (2003).

\bibitem{Ki08}
T. Kib\'{e}di, T. W. Burrows, M. B. Trzhaskovskaya, P. M. Davidson, and C. W. Nestor Jr., Nucl. Instrum. Meth.
in Phys. Res. {\bf A589}, 202 (2008).

\bibitem{Ba02} 
I. M. Band, M. B. Trzhaskovskaya, C. W. Nestor, Jr., P. Tikkanen, and S. Raman, 
At. Data Nucl. Data Tables {\bf 81}, 1 (2002).

\bibitem{So77}
S.K. Soni, A. Kumar, S.L. Gupta and S.C. Pancholi
Z. Physik A {\bf 282}, 49 (1977). 

\bibitem{Ka12}
A. Kahoul, V. Aylikci, N. Kup Aylikci, E. Cengiz and G. Apaydin,
Radiat. Phys. Chem. {\bf 81}, 713 (2012).

\bibitem{Sc96} 
E. Sch\"onfeld and H. Janssen, 
Nucl. Instrum. Methods Phys. Res. A {\bf 369}, 527 (1996).

\bibitem{Ha11}
A. Hashizume
Nuclear Data Sheets {\bf 112}, 1647 (2011)

\bibitem{Ha02}
J. C. Hardy, V. E. Iacob, M. Sanchez-Vega, R. T. Effinger, P. Lipnik, V. E. Mayes, D. K. Willis,
and R.G. Helmer, 
Appl. Radiat. Isot. {\bf 56}, 65 (2002).

\bibitem{He04}
R. G. Helmer, N. Nica, J. C. Hardy, and V. E. Iacob, 
Appl. Radiat. Isot. {\bf 60}, 173 (2004).

\bibitem{Ha92}
J. A. Halbleib, R. P. Kemsek, T. A. Melhorn, G. D. Valdez, S. M. Seltzer and M. J. Berger, Report SAND91-16734, Sandia National Labs (1992).

\bibitem{Rapc}
D. Radford, http://radware.phy.ornl.gov/main.html and private communication.

\bibitem{Fi96}
R.B. Firestone,\textit{Table of Isotopes}, ed. V.S. Shirley (John Wiley \& Sons Inc., New York, 1996) p F-44. 

\bibitem{Ch05}
C. T. Chantler, K. Olsen, R. A. Dragoset, J. Chang, A. R. Kishore, S. A. Kotochigova and D. S. Zucker (2005),
{\it X-Ray Form Factor, Attenuation and Scattering Tables (version 2.1)}. Available online at
http://physics.nist.gov/ffast.

\bibitem{Au65}
R.L. Auble and W.H. Kelly
Nucl. Phys. {\bf 73}, 25 (1965).

\bibitem{Ap70}
K.E. Apt, W.B. Walters and G.E. Gordon,
Nucl. Phys. A {\bf 152}, 344 (1970).



\end{thebibliography}
\end{document}